\documentclass[a4paper]{article}
\usepackage{RR}
\RRNo{6324}
\usepackage{hyperref}
\usepackage{amssymb}

\def\email#1{{\tt#1}}
\newtheorem{theorem}{Theorem}[section]{\bfseries}{\itshape}
\newtheorem{definition}{Definition}{\bfseries}{\itshape}
\newtheorem{proposition}{Proposition}{\bfseries}{\itshape}
\newtheorem{corollary}{Corollary}{\bfseries}{\itshape}
\newcommand{\proof}{\paragraph{\textsl{Proof: }}}

\RRdate{Octobre 2007}

\RRauthor{
Francis Klay\thanks[FK]{France Telecom R\&D, \email{francis.klay@orange-ftgroup.com}}%
  \and
Judson Santiago\thanks[JS]{DIMAp - UFRN, \email{judson@dimap.ufrn.br}}%
  \and
Laurent Vigneron\thanks[LV]{LORIA - Nancy Universit\'e, \email{laurent.vigneron@loria.fr}}%
}
\authorhead{Klay, Santiago \& Vigneron}
\RRtitle{M\'ethodes automatiques\\
  pour l'analyse de protocoles de non r\'epudiation\\
  en pr\'esence d'un intrus actif}
\RRetitle{Automatic Methods\\
  for Analyzing Non-Repudiation Protocols\\
  with an Active Intruder}
\titlehead{Analyzing Non-Repudiation Protocols}
\RRnote{This work is supported by the ANR AVOT\'E.}
\RRresume{%
  Les protocoles de non r\'epudiation ont un r\^ole important dans de
  nombreux domaines, o\`u des transactions s\'ecuris\'ees avec preuves
  de participation sont n\'ecessaires.  Les m\'ethodes formelles sont
  astucieuses et sans erreur, il est donc crucial de les utiliser pour
  v\'erifier de tels protocoles.  Dans ce but, nous montrons sur le
  protocole Fair Zhou-Gollmann comment repr\'esenter partiellement la
  non r\'epudiation comme la combinaison d'authentifications.  Apr\`es
  discussion des limites d'une telle approche, nous d\'efinissons une
  nouvelle m\'ethode bas\'ee sur la gestion des connaissances des
  participants du protocole.  Cette m\'ethode est tr\`es g\'en\'erale
  et d'usage naturel, car elle consiste \`a ajouter de simples
  annotations, comme pour les probl\`emes d'authentification.  La
  m\'ethode est tr\`es facile \`a implanter dans des outils capables
  de g\'erer les connaissances des participants.  Nous l'avons
  implant\'ee dans l'outil AVISPA et avons analys\'e le protocole
  optimiste Cederquist-Corin-Dashti, d\'ecouvrant deux attaques
  inconnues jusqu'alors.  Cette extension de l'outil AVISPA pour
  traiter la non r\'epudiation ouvre un boulevard pour sp\'ecifier de
  nombreuses autres propri\'et\'es, sans changement suppl\'ementaire
  dans l'outil.}
\RRabstract{%
  Non-repudiation protocols have an important role in many areas where
  secured transactions with proofs of participation are necessary.
  Formal methods are clever and without error, therefore using them
  for verifying such protocols is crucial.  In this purpose, we show
  how to partially represent non-repudiation as a combination of
  authentications on the Fair Zhou-Gollmann protocol.  After
  discussing its limits, we define a new method based on the handling
  of the knowledge of protocol participants.  This method is very
  general and is of natural use, as it consists in adding simple
  annotations, like for authentication problems.  The method is very
  easy to implement in tools able to handle participants knowledge.
  We have implemented it in the AVISPA Tool and analyzed the
  optimistic Cederquist-Corin-Dashti protocol, discovering two unknown
  attacks.  This extension of the AVISPA Tool for handling
  non-repudiation opens a highway to the specification of many other
  properties, without any more change in the tool itself.}
\RRmotcle{protocoles cryptographiques, non r\'epudiation, \'equit\'e,
  authentification, analyse automatique, outil AVISPA}
\RRkeyword{cryptographic protocols, non-repudiation,
  fairness, authentication, automatic analysis, AVISPA Tool}
\RRprojets{Cassis}
\RRtheme{\THSym}
\URLorraine 
\begin{document}
\makeRR   

\section{Introduction}

Considering security protocols, the study of properties such as
authentication and secrecy has been intensive for years~\cite{ryan00},
but the interest of other properties such as non-repudiation and
fairness has been raised only in the 1990s with the explosion of
Internet services and electronic transactions.%
\footnote{See
  \url{http://www.lsv.ens-cachan.fr/~kremer/FXbib/references.php} for
  a detailed list of publications related to the analysis of
  non-repudiation protocols.}

Non-repudiation protocols are designed for verifying that, when two
parties exchange information over a network, neither one nor the other
can deny having participated to this communication.  Such a protocol
must therefore generate evidences of participation to be used in case
of a dispute.  The basic tools for non-repudiation services have been
digital signatures and public key cryptography.  Indeed, when one
receives a signed message, he has an evidence of the participation and
the identity of his party~\cite{kremer02}.\\
The majority of the non-repudiation property analysis efforts in the
literature are manually driven though.  One of the first efforts to
apply formal methods to the verification of non-repudiation protocols
have been presented by Zhou et al. in~\cite{zhou98towards}, where they
used SVO logic.
In~\cite{schneider98} Schneider used process algebra CSP to prove the
correctness of a non-repudiation protocol, the well-known Fair
Zhou-Gollmann protocol.  With the same goal, Bella et al. have used
the theorem prover Isabelle~\cite{bella01}.  Schneider used a rank
function for encoding that in an execution trace, an event happens
before another event.  The verification is done by analyzing traces in
the stable failures models of CSP.
Among the automatic analysis attempts, we can cite Shmatikov and
Mitchell~\cite{shmatikov00analysis} who have used Mur$\varphi$, a
finite state model-checker, to analyze a fair exchange and two
contract signing protocols, Kremer and Raskin~\cite{kremer01gamebased}
who have used a game based model, Armando et
al.~\cite{ArmandoCC-CSF07} who used LTL for encoding resilient
channels in particular, the very nice work of Gurgens and
Rudolph~\cite{GurgensR-FAC05} who have used the asynchronous
product automata (APA) and the simple homomorphism verification tool
(SHVT)~\cite{SHVT-98}, raising flaws in three variants of
the Fair Zhou-Gollmann protocol and in two fair non-repudiation
protocols~\cite{KremerK-ITB00,ZhouDB-ACISP99}.
Wei and Heather~\cite{WeiH-FAST05} have used FDR, with an approach
similar to Schneider, for a variant of the Fair Zhou-Gollmann protocol
with timestamps.

The common point between all those works is that they use rich logics,
with a classical bad consequence for model checkers, the difficulty to
consider large protocols.
For avoiding this problem, Wei and Heather~\cite{WeiH-FAST06} used
PVS~\cite{Roscoe}, but some of the proof are still manual.\\

Fairness is more difficult to achieve: no party should be able to
reach a point where he has the evidence or the message he requires
without the other party also having his required evidence.  Fairness
is not always required for non-repudiation protocols, but it is
usually desirable.\\
A variety of protocols has been proposed in the literature to solve
the problem of fair message exchange with non-repudiation.  The first
solutions were based on a gradual exchange of the expected
information~\cite{kremer02}.  However this simultaneous secret
exchange is troublesome for actual implementations because fairness is
based on the assumption of equal computational power on both parties,
which is very unlikely in a real world scenario.  A possible solution
to this problem is the use of a trusted third party (TTP), and in fact
it has been shown that it is impossible to achieve fair exchange
without a TTP~\cite{pagnia99,markowitch99}.  The TTP can be used as a
delivery agent to provide simultaneous share of evidences.  The Fair
Zhou-Gollmann protocol~\cite{zhou96fair} is a well known example
using a TTP as a delivery agent;
a significant amount of work has been done over this protocol and its
derivations~\cite{bella01,gurgens03,schneider98,zhou98towards}.
However, instead of passing the complete message through the TTP and
thus creating a possible bottleneck, recent evolution of protocols
resulted in efficient, \emph{optimistic} versions, in which the TTP is
only involved in case anything goes wrong.  Resolve and abort
sub-protocols must guarantee that every party can complete the
protocol in a fair manner and without waiting for actions of the other
party.\\
One of these recent protocols is the optimistic
Cederquist-Corin-Dashti (CCD) non-repudia\-tion
protocol~\cite{cederquist05}.  The CCD protocol has the advantage of
not using session labels, contrariwise to many others in the
literature~\cite{kremer02,markowitch01,zhou96fair,schneider98}.  A
session label typically consists of a hash of all message components.
G{\"u}rgens et al.~\cite{gurgens03} have shown a number of
vulnerabilities associated to the use of session labels and, to our
knowledge, the CCD protocol is the only optimistic non-repudiation
protocol that avoids altogether the use of session labels.

This paper presents a method for automatically verifying
non-repudiation protocols in presence of an active intruder.  Our
method has been implemented in the AVISPA
Tool~\cite{AvispaCAV05}\footnote{\url{http://www.avispa-project.org}}
and we illustrate it with examples.  This tool, intensively used for
defining Internet security protocols and automatically analyzing their
authentication and secrecy properties, did not provide any help for
considering non-repudiation properties.\\
We first consider non-repudiation analysis as a combination of
authentication problems, applied to the Fair Zhou-Gollmann protocol.
We show the limits of this representation and the difficulties for
proving non-repudiation properties using only authentications.  Then,
we define method based on the analysis of agents knowledge, permitting
to handle non-repudiation and fairness properties in a same framework.
Our approach is very natural for the user and writing the logical
properties is still simple: they correspond to state invariants that
are convincing properties for the user.  This method is easy to
integrate in lazy verification systems, such as the AVISPA Tool, and
can also be integrated in any system able to handle agents (or
intruder) knowledge.  This should permit, contrarily to more complex
logics like LTL, to set up abstractions more easily for considering
unbounded cases.  This should also permit to get a more efficient
verification for bounded cases.
We illustrate this with the
optimistic Cederquist-Corin-Dashti protocol.

\section{Non-Repudiation Properties}\label{sec-prop-nr}

Non-repudiation (NR) is a general property that may not be clearly
defined.  It is usually described as a set of required services,
depending on the protocol and the required level of security.  In
particular, non-repudiation properties may be different whether a
trusted third party (TTP) is used or not in the protocol.

Considering a message sent by an originator agent to a recipient agent
(possibly via a delivery agent, a TTP), we define below some of the
most important non-repudiation services required by most of the
existing security applications (for e-commerce for example).
\begin{definition}
  The service of \textbf{non-repudiation of origin}, denoted ${\cal
    NRO}_B(A)$, provides the recipient $B$ with a set of evidences
  which ensures that the originator $A$ has sent the message.  The
  evidence of origin is generated by the originator and held by the
  recipient.  This property protects the recipient against a
  dishonest originator.
\end{definition}

\begin{definition}
  The service of \textbf{non-repudiation of receipt}, denoted ${\cal
    NRR}_A(B)$, provides the originator $A$ a set of evidences which
  ensures that the recipient $B$ has received the message.  The
  evidence of receipt is generated by the recipient and held by the
  originator.  This property protects the originator against a
  dishonest recipient.
\end{definition}

\begin{definition}
  The service of \textbf{non-repudiation of submission}, denoted
  ${\cal NRS}_A(B)$, provides the originator $A$ a set of evidences
  which ensures that he has submitted the message for delivery to $B$.
  This service only applies when the protocol uses a TTP.  Evidence of
  submission is generated by the delivery agent, and will be held by
  the originator.  This property protects the originator against a
  dishonest recipient.
\end{definition}

\begin{definition}
  The service of \textbf{non-repudiation of delivery}, denoted ${\cal
    NRD}_A(B)$, provides the originator $A$ a set of evidences which
  ensures that the recipient $B$ has received the message.  This
  service only applies when the protocol uses a TTP.  Evidence of
  delivery is generated by the delivery agent, and will be held by the
  originator.  This property protects the originator against a
  dishonest recipient.
\end{definition}

\begin{definition}
  A service of \textbf{fairness} (also called \textsl{strong
    fairness}) for a non-repudiation protocol provides evidences that
  if, at the end of the protocol execution, either the originator has
  the evidence of receipt of the message and the recipient has the
  evidence of origin of the corresponding message, or none of them has
  any valuable information.  This property protects the originator and
  the recipient.
\end{definition}

\begin{definition}
  A service of \textbf{timeliness} for a non-repudiation protocol
  guarantees that, whatever happens during the protocol run, all
  participants can reach a state that preserves fairness, in a finite
  time.
\end{definition}
Note that in general sets of evidences such as $\cal NRO$, $\cal NRR$,
$\cal NRS$ and $\cal NRD$ are composed with messages signed by an
agent.

For the sequel of this paper, we will consider the following definition
of an evidence.
\begin{definition}
  An \textbf{evidence} for an agent $A$ for a non-repudiation property
  $P$ is a message, a part of a message, or a combination of both,
  received by $A$ that is necessary for guaranteeing property $P$.
\end{definition}

Note that in this paper, we consider the evidences given by the
protocol designer as valid: without intervention of an intruder, those
evidences are sufficient to guarantee the non-repudiation service; and
in case of a dispute, a judge analyzing them will always be able to
protect honest agents.

\section{Non-Repudiation as Authentication}

It is well known that non-repudiation is a form of
authentication~\cite{ryan00}.  In this section we demonstrate that
properties like $\cal NRO$, $\cal NRR$,\ldots can be at least
partially represented by authentication properties.  We illustrate
this idea with the Fair Zhou-Gollmann protocol.  At the end of this
section we show strong limitations of this approach in order to
motivate the introduction of a new approach in the next section.

\subsection{Running Example: the FairZG Protocol}

In this section we describe the Fair Zhou-Gollmann protocol
(FairZG)~\cite{zhou98towards}, a fair non-repudiation protocol that
uses a TTP.  We have chosen this protocol as a case study to
demonstrate our analysis approach because of the existence of
significant related work~\cite{bella01,gurgens03,schneider98}.
The protocol is presented below in Alice\&Bob notation, where
\textsf{fNRO}, \textsf{fNRR}, \textsf{fSUB} and \textsf{fCON} are
labels used to identify the purpose of messages.
\begin{tabbing}
  xxxx \= xxxxxxxxx              \= \kill
  ~~1.  \> {\sf A $\rightarrow$ B:}       \> {\sf fNRO.B.L.C.NRO}\\
  ~~2.  \> {\sf B $\rightarrow$ A:}       \> {\sf fNRR.A.L.NRR}\\
  ~~3.  \> {\sf A $\rightarrow$ TTP:}     \> {\sf fSUB.B.L.K.SubK}\\
  ~~4.  \> {\sf B $\leftrightarrow$ TTP:} \> {\sf fCON.A.B.L.K.ConK}\\
  ~~5.  \> {\sf A $\leftrightarrow$ TTP:} \> {\sf fCON.A.B.L.K.ConK}\\[1mm]
  and \>
  ${\cal NRO}_B(A) = \{ \mathsf{NRO}, \mathsf{ConK} \}$\\
  \> ${\cal NRR}_A(B) = \{ \mathsf{NRR}, \mathsf{ConK} \}$
\end{tabbing}
where
\textsf{A} (for Alice) is the originator of the message \textsf{M},
\textsf{B} (for Bob) is the recipient of the message \textsf{M},
\textsf{TTP} is the trusted third party,
\textsf{M} is the message to be sent from Alice to Bob,
\textsf{C} is a commitment (the message \textsf{M} encrypted by a key
  \textsf{K}),
\textsf{L} is a unique session identifier (also called label),
\textsf{K} is a symmetric key defined by Alice,
\textsf{NRO} is a message used for non-repudiation of origin (the message
  \textsf{fNRO.B.L.C} signed by Alice),
\textsf{NRR} is a message used for non-repudiation of receipt (the message
  \textsf{fNRR.A.L.C} signed by Bob),
\textsf{SubK} is a proof of submission of \textsf{K} (the message
  \textsf{fSUB.B.L.K} signed by A),
\textsf{ConK} is a confirmation of \textsf{K} (the message
  \textsf{fCON.A.B.L.K} signed by the TTP).

The main idea of the FairZG protocol is to split the delivery of a
message into two parts.  First a commitment \textsf{C}, containing the
message \textsf{M} encrypted by a key \textsf{K}, is exchanged between
Alice and Bob (message \textsf{fNRO}).  Once Alice has an evidence of
commitment from Bob (message \textsf{fNRR}), the key \textsf{K} is
sent to a trusted third party (message \textsf{fSUB}).  Once the TTP
has received the key, both Alice and Bob can retrieve the evidence
\textsf{ConK} and the key \textsf{K} from the TTP (messages
\textsf{fCON}).  This last step is represented by a double direction
arrow in the Alice\&Bob notation because it is implementation specific
and may be composed by several message exchanges between the agents
and the TTP.  In this scenario we assume the network will not be down
forever and both Alice and Bob have access to the TTP's shared
repository where it stores the evidences and the key.  This means that
the agents will eventually be able to retrieve the key and evidences
from the TTP even in case of network failures.

\subsection{Non-Repudiation of Origin as Authentication}\label{ssec:nro_as_auth} 
In our example, the FairZG protocol, non-repudiation of origin should
provide the guarantee that if Bob owns ${\cal NRO}$ then Alice has
sent \textsf{M} to Bob.  Proposition~\ref{prop:auth_nro} shows how
this can be partially done with a set of authentications.

\begin{definition}
  \textsf{auth(X,Y,D)} is the non-injective authentication, and means
  \textsf{X} authenticates \textsf{Y} on data \textsf{D}.
\end{definition}
The semantics of such a predicate is standard and can be found
in~\cite{lowe-csfw97}.

\begin{proposition}\label{prop:auth_nro}
  Given the FairZG protocol, let \textsf{B} be a honest agent.\\
  If \textsf{auth(B,A,NRO)}, \textsf{auth(B,TTP,ConK)} and
  \textsf{auth(TTP,A,SubK)} are satisfied then the non-repudiation
  service of origin ${\cal NRO}_B(A)$ is satisfied.
\end{proposition}

\proof{
  For the two evidences of ${\cal NRO}_B(A) = \{ \mathsf{NRO},
  \mathsf{ConK} \}$, we have:
  \begin{itemize}
  \item $\mathsf{NRO = Sig_A(fNRO.B.L.\{M\}_K)}$: since
    \textsf{auth(B,A,NRO)} is satisfied, there is an agreement on
    $\mathsf{Sig_A(fNRO.B.L.C)}$ between \textsf{B} and \textsf{A}.
    From the signature properties this means also an agreement on
    $\mathsf{\{M\}_K}$, thus \textsf{A} has sent
    $\mathsf{\{M\}_K}$.
  \item $\mathsf{ConK = Sig_{TTP}(fCON.A.B.L.K)}$: as above
    \textsf{auth(B,TTP,ConK)} implies an agreement on \textsf{K}
    between \textsf{B} and \textsf{TTP}.  Furthermore $\mathsf{SubK =
      Sig_A(fSUB,B,L,K)}$ thus \textsf{auth(TTP,A,SubK)} implies an
    agreement on \textsf{K} between \textsf{TTP} and \textsf{A}.  By
    transitivity we have an agreement on \textsf{K} between \textsf{B}
    and \textsf{A} which means that \textsf{A} has sent \textsf{K}.
  \end{itemize}
  As \textsf{A} has sent $\mathsf{\{M\}_K}$ and \textsf{K}, he has
  sent \textsf{M}.  The non-injective authentication is only required
  for \textsf{auth(B,TTP,ConK)} because \textsf{B} can ask many times
  \textsf{ConK}.  However since all authentications imply an agreement
  on the unique session identifier \textsf{L}, this excludes an
  authentication across different sessions.
\hfill$\Box$}

\subsection{Non-Repudiation of Receipt as Authentication}\label{ssec:nrr_as_auth} 
In our example, the FairZG protocol, non-repudiation of receipt should
provide the guarantee that if Alice owns ${\cal NRR}$ then Bob has
receipt \textsf{M} from Alice.  Proposition~\ref{prop:auth_nrr} shows
how this can be done partially with a set of authentications.

\begin{proposition}\label{prop:auth_nrr}
  Given the FairZG protocol, let \textsf{B} be a honest agent.\\
  If \textsf{auth(A,B,NRR)}, \textsf{auth(A,TTP,ConK)} and
  \textsf{auth(B,TTP,ConK)} are satisfied then the non-repudiation
  service of receipt ${\cal NRR}_A(B)$ is satisfied.
\end{proposition}

\proof{
  For the two evidences of ${\cal NRR}_A(B) = \{ \mathsf{NRR},
  \mathsf{ConK} \}$, we have:
  \begin{itemize}
  \item $\mathsf{NRR = Sig_B(fNRR.A.L.\{M\}_K)}$: a reasoning as for
    \textsf{NRO} in Proposition~\ref{prop:auth_nro} ensures that
    \textsf{B} has received $\mathsf{\{M\}_K}$.
  \item $\mathsf{ConK = Sig_{TTP}(fCON.A.B.L.K)}$:
    \textsf{auth(A,TTP,ConK)} implies an agreement on \textsf{K}
    between \textsf{A} and \textsf{TTP}.  Furthermore
    \textsf{auth(B,TTP,ConK)} implies an agreement on \textsf{K}
    between \textsf{B} and \textsf{TTP}. This means that there is an
    agreement on \textsf{K} between \textsf{A} and \textsf{B}, thus
    when \textsf{A} holds \textsf{ConK}, \textsf{B} has received or
    will be able to receive \textsf{K}.
  \end{itemize}
  The proof end is similar to the one of
  Proposition~\ref{prop:auth_nro}.
\hfill$\Box$}

\subsection{Limitations and Difficulties}

At this point there are some problems that motivate the introduction
of a new approach presented in the next section.
\begin{enumerate}
\item If, contrarily to the previous Propositions hypothesis, the
  evidences owner is dishonest, he can possibly forge a fake evidences
  set.  For example for Bob and ${\cal NRR}$ we need to prove that Bob
  could only own ${\cal NRR}$ if Alice has actually sent the correct
  protocol messages.  This may be done as for example
  in~\cite{schneider98}, \cite{WeiH-FAST05} or \cite{gurgens03} but
  this is not trivial.
\item Handling non-repudiation as authentications seems very hard or
  may not be possible in general.  In particular this task seems
  difficult for optimistic non-repudiation protocols that include
  sub-protocols like \textsl{abort} and \textsl{resolve} as presented
  in the next section.
\item In general verifying Fairness is a delicate stage and the above
  remarks make this more difficult.
\end{enumerate}

In conclusion, proving non-repudiation with the help of
authentications seems for us not to be the right way; this is why in
the next section we propose a very easy approach for handling
non-repudiation.

\section{Non-Repudiation based on Agent Knowledge}

In this section, we present a new method for considering
non-repudiation services and fairness in a same framework: we
introduce a logic permitting to describe states invariants.  This
logic is a very classical one, except that we define two new
predicates, \texttt{deduce} and \texttt{aknows} that permit to
consider agents knowledge in the description of goals.  The
\texttt{aknows} predicate is also used as protocol annotation, with
the semantics \textit{agent $X$ knows (or can deduce) term $t$}.

\subsection{Description of Non-Repudiation Properties}
\label{sec-descrNR}

The main role of a non-repudiation protocol is to give evidences of
non-repudiation to the parties involved in the protocol.  To analyze
this kind of protocol, one must verify which participants have their
non-repudiation evidences at the end of the protocol execution.  For
example, if the originator has all its evidences for non-repudiation
of receipt, then the service of non-repudiation of receipt is
guaranteed.  If the recipient has all its evidences for
non-repudiation of origin, then the service of non-repudiation of
origin is guaranteed.  If both parties (or none of them) have their
evidences, fairness is guaranteed.  In other words, to analyze
non-repudiation, we need to verify if a set of terms is known by an
agent at the end of the protocol execution.

And for considering a large class of non-repudiation protocols, we
shall not restrict evidences to a set of terms, but we have to
consider them as a combination of terms using standard logical
connectors (conjunction, disjunction, negation).

For considering non-repudiation and fairness properties involving
honest and dishonest agents, we have defined a new predicate that
permits to access the knowledge of protocol participants.  This
predicate, named \texttt{aknows}, is used in the specification of
protocol transitions and of properties.

\begin{definition}[$\mathcal{NR\_}_{X}(Y)$]
  Let $\mathcal{A}$ be a set of agents playing a finite number of
  sessions $\mathcal{S}$ of a protocol, $\mathcal{T}$ a set of terms
  sent in the messages of this protocol and $\mathcal{E}$ the subset
  of terms in $\mathcal{T}$ that are part of the evidences of
  non-repudiation in the protocol.  For an agent $X \in \mathcal{A}$,
  $\mathcal{NR\_}_{X}(Y)$ is a logical combination of terms $t \in
  \mathcal{E}$ that constitute the evidence for a service of
  non-repudiation $\cal NR\_$ for agent $X$ wrt.\ agent $Y$.
\end{definition}

\begin{definition}[\texttt{aknows}]
  Let $\mathcal{A}$ be a set of agents playing a finite number of
  sessions $\mathcal{S}$ of a protocol, $\mathcal{T}$ a the set of
  terms. The annotation $\mathtt{aknows}(X,s,t)$ is a predicate with
  $X\in\mathcal{A}$, $s \in \mathcal{S}$ and $t \in \mathcal{T}$,
  expressing that agent $X$, playing in session $s$ of the protocol,
  knows (or can deduce) the term $t$.
\end{definition}
The semantics of predicate $\mathtt{aknows}(X,s,t)$ is that the term
$t$ can be composed by agent $X$, according to its current knowledge
in the session $s$ of the protocol, whether this agent is honest or
not.  This composability test can be easily done by any tool that is
able to manage agents knowledge or intruder knowledge.

By abuse of notation, we may write $\mathtt{aknows}(X,s,L)$, for a
logical formula $L$ combining evidences ($\mathcal{NR\_}_{X}(Y)$ for
example), considering that the predicate $\mathtt{aknows}$ is an
homomorphism:
\begin{eqnarray*}
  \mathtt{aknows}(X,s,L_1 \wedge L_2) & = & \mathtt{aknows}(X,s,L_1)
  \wedge \mathtt{aknows}(X,s,L_2)\\
  \mathtt{aknows}(X,s,L_1 \vee L_2) & = & \mathtt{aknows}(X,s,L_1)
  \vee \mathtt{aknows}(X,s,L_2)\\
  \mathtt{aknows}(X,s,\neg L) & = & \neg \mathtt{aknows}(X,s,L)
\end{eqnarray*}

\begin{definition}[\texttt{deduce}]
  Let $\mathcal{A}$ be a set of agents playing a finite number of
  sessions of a protocol and $\mathcal{T}$ a set of terms.  We define
  $\mathtt{deduce}(X,t)$, with $X\in\mathcal{A}$ and $t \in
  \mathcal{T}$, as the predicate which means that \texttt{X} can
  deduce \texttt{t} from its knowledge.
\end{definition}
We will use the same abuse of notation for $\mathtt{deduce}$ as for
$\mathtt{aknows}$.\\

In the following, we assume that each \texttt{aknows} annotation
corresponds to a valid \texttt{deduce} predicate on the same
information, in order to avoid bad annotations.

\begin{definition}
  The evidence ${\cal NR\_}_X(Y)$ is \textbf{well-formed} if it
  contains information that uniquely identifies the session, and if it
  contains an injective function of the message $M$ for which ${\cal
    NR\_}$ acts as a protection agains a dishonnest agent.
\end{definition}

We now give the results obtained by this representation.
\begin{proposition}
  Given a non-repudiation service of $B$ against $A$ about a message $M$ with
  the well-formed evidence ${\cal NR\_}_B(A)$ in session $s$ of a protocol.
  If the following formulae are true at the session end then the non
  repudiation service is valid.
  \[
  \begin{array}{lcl}
  \mathtt{aknows}(B,s,{\cal NR\_}_B(A)) 
  &\Rightarrow 
  &\mathtt{aknows}(A,s,M)
  \\
  \mathtt{deduce}(B,{\cal NR\_}_B(A))   
  &\Rightarrow 
  &\mathtt{aknows}(B,s,{\cal NR\_}_B(A))\\
  \end{array}
  \]
\end{proposition}
\proof{A sketch of proof is as follows: by the second implication if
  $B$ is able to deduce ${\cal NR\_}_B(A)$ then
  $\mathtt{aknows}(B,s,{\cal NR\_}_B(A))$ is included in its
  knowledge.  Furthermore since ${\cal NR\_}_B(A)$ is well-formed,
  ${\cal NR\_}_B(A)$ and $\mathtt{aknows}(B,s,{\cal NR\_}_B(A))$ are
  related to the same session.

  Now since ${\cal NR\_}_B(A)$ is well-formed it includes all the
  information in $M$, thus the first implication implies an agreement
  on $M$ between $B$ and $A$.  Finally as $\mathtt{aknows}(A,s,M)$ is
  an annotation, this means that $A$ has followed the protocol, thus
  he has done what he must do with $M$.
\hfill$\Box$}

\noindent
\textit{Remark: } verifying formulas given in the above Proposition
is not a problem, because a priori any theorem prover can compute
whatever can be deduced by an agent at a given step of the protocol,
especially concerning the \texttt{deduce} predicate.

\begin{corollary}\label{cor-nro}
  Given a non-repudiation service of origin for $B$ against $A$ about
  message $M$, in session $s$ of a protocol. If ${\cal NRO}_B(A))$ is
  well-formed and the following formulae are true at the session end
  then the service is valid.
  \[
  \begin{array}{lcl}
  \mathtt{aknows}(B,s,{\cal NRO}_B(A)) 
  &\Rightarrow 
  &\mathtt{aknows}(A,s,M)
  \\
  \mathtt{deduce}(B,{\cal NRO}_B(A))   
  &\Rightarrow 
  &\mathtt{aknows}(B,s,{\cal NRO}_B(A))\\
  \end{array}
  \]
\end{corollary}

\begin{corollary}\label{cor-nrr}
  Given a non-repudiation service of receipt for $A$ against $B$ about
  message $M$, in session $s$ of a protocol. If ${\cal NRR}_A(B))$ is
  well-formed and the following formulae are true at the session end
  then the service is valid.
  \[
  \begin{array}{lcl}
  \mathtt{aknows}(A,s,{\cal NRR}_A(B)) 
  &\Rightarrow 
  &\mathtt{aknows}(B,s,M)
  \\
  \mathtt{deduce}(A,{\cal NRR}_A(B))   
  &\Rightarrow 
  &\mathtt{aknows}(A,s,{\cal NRR}_A(B))\\
  \end{array}
  \]
\end{corollary}

\subsection{Description of Fairness}

In the literature, authors often give different definitions of
fairness for non-repudiation protocols.  In some definitions none of
the parties should have more evidences than the others at any given
point in time.  Others have a more flexible definition in which none
of them should have more evidences than the others in the end of the
protocol run.  In many works it is also not very clear if only
successful protocol runs are taken into account, or partial protocol
runs are valid as well.

In this paper the later definition of fairness will be used and we
take into account complete protocol runs.  By complete protocol runs
we mean a run where, even though the protocol could not have reached
it's last transition for all agents, there is no executable transition
left, i.e.\ all possible protocol steps have been executed, but this
does not mean that all agents are in a final state.

We define this standard fairness as a function of non-repudiation of
origin and of non-repudiation of receipt.  If both properties, $\cal
NRO$ and $\cal NRR$, are ensured or both are not satisfied for a given
message $M$, then we have fairness.
\begin{proposition}\label{prop-f-auth}
  Given a protocol whose purpose is to send a message from Alice to
  Bob, we have the following equivalence concerning the standard
  definition of fairness for a given session $s$.
  If the non-repudiation is valid for the $\mathtt{{\cal
  NRO}}$ and $\mathtt{{\cal NRR}}$ services then:
  \[
  \mbox{Fairness} ~\equiv~
  \mathtt{aknows}(Bob,s,{\cal NRO}_{\mbox{Bob}}(\mbox{Alice}))
  \mbox{ iff }
  \mathtt{aknows}(Alice,s,{\cal NRR}_{\mbox{Alice}}(\mbox{Bob}))
  \]
\end{proposition}

This result can be generalized to fairness wrt.\ a set of
non-repudiation services as follows.
\begin{theorem}\label{th-f-auth}
  Given a protocol involving a finite number of agents,
  given a finite set of valid non-repudiation services $\cal NR$,
  the protocol is fair wrt.\ $\cal NR$ iff
  \[\begin{array}{l}
    \forall {{\cal NR}S_1}_{X_1}(Y_1), {{\cal NR}S_2}_{X_2}(Y_2) \in
    {\cal NR},~\\
    \hspace*{2cm}
    \mathtt{aknows}(X_1,s,{{\cal NR}S_1}_{X_1}(Y_1))
    ~\mbox{ iff }~
    \mathtt{aknows}(X_2,s,{{\cal NR}S_2}_{X_2}(Y_2))
  \end{array}\]
\end{theorem}

\subsection{Running Example: CCD}

For illustrating the analysis method described later on, we will use a
recent protocol, the optimistic Cederquist-Corin-Dashti (CCD)
non-repudiation protocol~\cite{cederquist05}.  The CCD protocol has
been created for permitting an agent $A$ to send a message $M$ to an
agent $B$ in a fair manner.  This means that agent $A$ should get an
evidence of receipt of $M$ by $B$ ($EOR$) if and only if $B$ has
really received $M$ and the evidence of origin from $A$ ($EOO$).
$EOR$ permits $A$ to prove that $B$ has received $M$, while $EOO$
permits $B$ to prove that $M$ has been sent by $A$.  The protocol is
divided into three sub-protocols: the main protocol, an \textsl{abort}
sub-protocol and a \textsl{resolve} sub-protocol.

\paragraph{The Main Protocol.}
It describes the sending of $M$ by $A$ to $B$ and the exchange of
evidences in the case where both agents can complete the entire
protocol.  If a problem happens to one of the agents, in order to
finish properly the protocol, the agents execute the \textsl{abort} or
the \textsl{resolve} sub-protocol with a trusted third party ($TTP$).

The main protocol is therefore composed of the following messages
exchanges, described in the Alice\&Bob notation:\\[2mm]
\begin{tabular}{l@{\hspace{0.2cm}}l@{\hspace{0.2cm}}l@{\hspace{0.5cm}}l}
  \small{1.} & $A \rightarrow B:$ & ${\{M\}}_{K}$.${EOO}_{M}$ 
  & where ${EOO}_{M} = {\{B.TTP.H({\{M\}}_{K}).{\{K.A\}}_{Kttp}\}}_{inv(Ka)}$\\
  \small{2.} & $B \rightarrow A:$ & ${EOR}_{M}$                      
  & where ${EOR}_{M} = {\{{EOO}_{M}\}}_{inv(Kb)}$\\
  \small{3.} & $A \rightarrow B:$ & $K$\\
  \small{4.} & $B \rightarrow A:$ & ${EOR}_{K}$                      
  & where ${EOR}_{K} = {\{A.H({\{M\}}_{K}).K\}}_{inv(Kb)}$\\
\end{tabular}\\[2mm]
where $K$ is a symmetric key freshly generated by $A$, $H$ is a
one-way hash function, $Kg$ is the public key of agent $g$ and
$inv(Kg)$ is the private key of agent $g$ (used for signing messages).
Note that we assure that all public keys are known by all agents
(including dishonest agents).

In the first message, $A$ sends the message $M$ encrypted by $K$ and
the evidence of origin for $B$ (message signed by $A$, so decryptable
by $B$).  In this evidence, $B$ can check his identity, learns the
name of the TTP, can check that the hash code is the result of hashing
the first part of the message, but cannot decrypt the last part of the
evidence; this last part may be useful if any of the other
sub-protocols is used.\\
$B$ answers by sending the evidence of receipt for $A$, $A$ checking
that $EOR_M$ is $EOO_M$ signed by $B$.\\
In the third message, $A$ sends the key $K$, permitting $B$ to
discover the message $M$.\\
Finally, $B$ sends to $A$ another evidence of receipt, permitting $A$
to check that the symmetric key has been received by $B$.

\paragraph{The \textsl{Abort} Sub-Protocol.}
The \textsl{abort} sub-protocol is executed by agent $A$ in case he
does not receive the message ${\mbox{EOR}}_{M}$ at step 2 of the main
protocol.  The purpose of this sub-protocol is to cancel the messages
exchange.
\begin{center}
  \begin{tabular}{l@{\hspace{0.2cm}}l@{\hspace{0.2cm}}l@{\hspace{0.5cm}}l}
    \small{1.} & $A \rightarrow TTP:$ &  ${\{\texttt{abort}.H({\{M\}}_{K}).B.{\{K.A\}}_{Kttp}\}}_{inv(Ka)}$\\
    \small{2.} & $TTP \rightarrow A:$ & 
    $\left\{
      \begin{array}{ll}
        {E}_{TTP}  & \mbox{ where } {E}_{TTP} = {\{A.B.K.H(\{M\}_{K})\}}_{inv(Kttp)}\\
        & \mbox{ if } \texttt{resolved}(A.B.K.H(\{M\}_{K}))\\
        {AB}_{TTP} & \mbox{ where } {AB}_{TTP} = {\{A.B.H(\{M\}_{K}).{\{K.A\}}_{Kttp}\}}_{inv(Kttp)}\\
        & \mbox{ otherwise}\\
      \end{array}
    \right.$\\
\end{tabular}
\end{center}
In this sub-protocol, $A$ sends to the TTP an abort request,
containing the \texttt{abort} label and some information about the
protocol session to be aborted.\\
According to what happened before, the TTP has two possible answers:
if this is the first problem received by the TTP for this protocol
session, the TTP sends a confirmation of abortion, and stores in its
database that this protocol session has been aborted; but if the TTP
has already received a request for resolving this protocol session, he
sends to $A$ the information for completing his evidence of receipt by
$B$.

\paragraph{The \textsl{Resolve} Sub-Protocol.}
The role of this second sub-protocol is to permit agents $A$ and $B$
to finish the protocol in a fair manner, if the main protocol cannot
be run until its end by some of the parties.  For example, if $B$
does not get $K$ or if $A$ does not get $EOR_K$, they can invoke the
\textsl{resolve} sub-protocol.
\begin{center}
  \begin{tabular}{l@{\hspace{0.2cm}}l@{\hspace{0.2cm}}l@{\hspace{0.5cm}}l}
    \small{1.} & $G \rightarrow TTP:$ &  ${EOR}_{M}$\\
    \small{2.} & $TTP \rightarrow G:$ & 
    $\left\{
      \begin{array}{ll}
        {AB}_{TTP}  & \mbox{ if } \texttt{aborted}(A.B.K.H(\{M\}_{K}))\\
        {E}_{TTP}   & \mbox{ otherwise}\\
      \end{array}
    \right.$
  \end{tabular}
\end{center}
where $G$ stands for $A$ or $B$.

A resolve request is done by sending ${EOR}_{M}$ to the TTP.  If the
protocol session has already been aborted, the TTP answers by the
abortion confirmation.  If this is not the case, the TTP sends
$E_{TTP}$ so that the user could complete its evidence of receipt (if
$G$ is $A$) or of origin (if $G$ is $B$).
Then the TTP stores in its database that this protocol session has
been resolved.

\paragraph{Agents' Evidences.}
For this protocol, according to~\cite{cederquist05}, the logical
formulas of evidences are:
\[\begin{array}{l}
  {\cal NRO}_B(A) = \{M\}_K \wedge EOO_M \wedge K\\
  {\cal NRR}_A(B) = \{M\}_K \wedge EOR_M \wedge (EOR_K \vee E_{TTP})
\end{array}\]
Note that there are two possibilities of evidences for non-repudiation
of receipt, according to the way the protocol is run.\\

According to our method, we simply have to annotate protocol steps
with \texttt{aknows} predicates, and then write the logical formula to
verify.  The following table shows where those annotations take place
in the three CCD sub-protocols, for considering non-repudiation of
origin and of receipt.
\begin{center}
\begin{tabular}[t]{|c|c|}\hline
  ${\cal NRO}_B(A)$ & Protocol - step\\\hline\hline
  $\mathtt{aknows}(B,s,\{M\}_K)$ & Main - 1.\\\hline
  $\mathtt{aknows}(B,s,EOO_M)$ & Main - 1.\\\hline
  $\mathtt{aknows}(B,s,K)$ & Main - 3.\\\hline
  $\mathtt{aknows}(B,s,K)$ & Resolve - 2.\\\hline
\end{tabular}
\hspace*{5mm}
\begin{tabular}[t]{|c|c|}\hline
  ${\cal NRR}_A(B)$ & Protocol - step\\\hline\hline
  $\mathtt{aknows}(A,s,\{M\}_K)$ & Main - 1.\\\hline
  $\mathtt{aknows}(A,s,EOR_M)$ & Main - 2.\\\hline
  $\mathtt{aknows}(A,s,EOR_K)$ & Main - 4.\\\hline
  $\mathtt{aknows}(A,s,E_{TTP})$ & Abort - 2.\\\hline
  $\mathtt{aknows}(A,s,E_{TTP})$ & Resolve - 2.\\\hline
\end{tabular}
\end{center}

According to Corollary~\ref{cor-nro}, \textbf{non-repudiation of
  origin} for the CCD protocol is represented by the following
invariant formulas:
\[\begin{array}{l}
  \mathtt{aknows}(B,s,\{M\}_K
  \wedge EOO_M
  \wedge K
  \Rightarrow
  \mathtt{aknows}(A,s,M)\\
  \mathtt{deduce}(B,\{M\}_K
  \wedge EOO_M
  \wedge K)
  \Rightarrow
  \mathtt{aknows}(B,s,\{M\}_K
  \wedge EOO_M
  \wedge K)
\end{array}\]

According to Corollary~\ref{cor-nrr}, \textbf{non-repudiation of
  receipt} for the CCD protocol is represented by the following
invariant formulas:
\[\begin{array}{l}
  \mathtt{aknows}(A,s,\{M\}_K
  \wedge EOR_M
  \wedge (EOR_K \vee E_{TTP}))
  \Rightarrow
  \mathtt{aknows}(B,s,M)\\
  \mathtt{deduce}(A,s,\{M\}_K
  \wedge EOR_M
  \wedge (EOR_K \vee E_{TTP}))\\
  \hspace*{3cm}\Rightarrow
  \mathtt{aknows}(A,s,\{M\}_K
  \wedge EOR_M
  \wedge (EOR_K \vee E_{TTP}))
\end{array}\]

For analyzing \textbf{fairness}, this protocol requires timeliness,
that is each participant should reach a final state before testing
fairness.
Fairness for the CCD protocol is described by the following logical
formulas, a very simple application of Theorem~\ref{th-f-auth}:
\[
\mathtt{aknows}(A,s,{\cal NRR}_A(B))
\Leftrightarrow
\mathtt{aknows}(B,s,{\cal NRO}_B(A))
\]
Basically the property states that if $A$ knows the EOR evidence
(${\{M\}}_{K}$, ${EOR}_{M}$, and ${EOR}_{K}$ or ${E}_{TTP}$), then $B$
must know the EOO evidence.  And symmetrically for $B$, if $B$ knows
the EOO evidence (${\{M\}}_{K}$, ${EOO}_{M}$, and $K$ or ${E}_{TTP}$),
then $A$ must know the EOR evidence.\\

The CCD protocol has been specified in the AVISPA Tool, with the
description of the fairness property given above.  The detailed
formulas used in the AVISPA Tool, with an LTL syntax, are:
\begin{small}\[
\Box \left( \left( \begin{array}{lll}
      \mathtt{aknows}(A,s,{\{M\}}_{K}) \; \wedge\\
      \mathtt{aknows}(A,s,{EOR}_{M}) \; \wedge\\
      (\mathtt{aknows}(A,s,{EOR}_{K}) \vee \mathtt{aknows}(A,s,{E}_{TTP}))\\
    \end{array} \right)
  \Rightarrow
  \left( \begin{array}{ll}
      \mathtt{aknows}(B,s,{\{M\}}_{K}) \; \wedge\\
      \mathtt{aknows}(B,s,{EOO}_{M}) \; \wedge\\
      \mathtt{aknows}(B,s,K)\\
    \end{array} \right) \right)
\]\end{small}%
\begin{small}\[
\Box \left( \left( \begin{array}{lll}
      \mathtt{aknows}(B,s,{\{M\}}_{K}) \; \wedge\\
      \mathtt{aknows}(B,s,{EOO}_{M}) \; \wedge\\
      \mathtt{aknows}(B,s,K)\\
    \end{array} \right)
  \Rightarrow
  \left( \begin{array}{ll}
      \mathtt{aknows}(A,s,{\{M\}}_{K}) \; \wedge\\
      \mathtt{aknows}(A,s,{EOR}_{M}) \; \wedge\\
      (\mathtt{aknows}(A,s,{EOR}_{K}) \vee \mathtt{aknows}(A,s,{E}_{TTP}))\\
    \end{array} \right) \right)
\]\end{small}%

Several scenarios have been run, and two of them have raised an
attack, showing that the CCD protocol does not provide the fairness
property for which it has been designed.

The first attack has been found for a scenario where only one session
of the protocol is run, between honest agents.  The problem is raised
when some messages of the main protocol are delayed, either by a slow
network traffic or by the action of an intruder.  The consequence of
this delay is that $A$ will invoke the \textsl{abort} sub-protocol and
$B$ will invoke the \textsl{resolve} sub-protocol.  And if the resolve
request reaches the TTP before the abort request, $B$ will get all his
necessary evidences from the TTP, while $A$ is not able to get all his
evidences even with the help of the TTP.\\
The originality of this attack is that, at the end:
\begin{itemize}
\item $A$ will guess (according to the answer received to his abort
  request) that the protocol has been resolved by $B$, so he will
  assume that $B$ knows $M$ and can build the proof that $A$ has sent
  it; but $A$ cannot prove this;
\item $B$ has resolved the protocol and has received from the TTP the
  information for getting $M$ and building the proof that $A$ has sent
  $M$; but he does not know that $A$ does not have his proof;
\item the TTP will think that $B$ has asked for the protocol to be
  resolved, followed by $A$; so for him, both $A$ and $B$ can build
  their evidences.
\end{itemize}
So, this trace shows that the CCD protocol is not fair, even if both
agents $A$ and $B$ are honest.  The attack is due to a malicious
intruder or a network problem, and the TTP is of no help for detecting
the problem.

The second attack is a variant: it happens when agent $A$ plays the
protocol with a dishonest agent $B$ (named $i$, for
\textsl{intruder}).  As soon as $i$ has received the first message
from $A$, he builds $EOR_{M}$ and sends it to the TTP as resolve
request.  When $A$ decides to abort the protocol, this is too late:
the protocol has already been resolved, the intruder can get $M$ and
build the proof that $A$ has sent $M$, and $A$ cannot build the
evidence of receipt.
We have corrected the protocol and the numerous scenarios that have
been tried on the new version have not raised any attack.
This experiment on the CCD protocol is detailed
in~\cite{SantiagoV-WISTP07}.

\section{Conclusion}

Non-repudiation protocols have an important role in many areas where
secured transactions with proofs of participation are necessary.  The
evidences of origin and receipt of a message are two examples of
elements that the parties should have at the end of the communication.
We have given two very different examples of such protocols.  The
FairZG protocol is an intensively studied protocol in which the role
of the trusted third party is essential.  The CCD protocol is a more
recent non-repudiation protocol that avoids the use of session labels
and distinguishes itself by the use of an optimistic approach, the
trusted third party being used only in case of a problem in the
execution of the main protocol.

The fairness of a non-repudiation protocol is a property difficult to
analyze and there are very few tools that can handle the automatic
analysis of this property.  The contribution of this work is twofold.
First, we have illustrated with the FairZG protocol how difficult it
is to consider full non-repudiation properties using only a
combination of authentications.

Second, we have defined a new method that permits to handle in a very
easy way non-repudiation properties and fairness in a same framework.
This method is based on the handling of agents knowledge and can be
used to automatically analyze non-repudiation protocols as well as
contract signing protocols~\cite{shmatikov00analysis}.  We have
implemented it in the AVISPA Tool and have successfully applied it to
the CCD protocol, proving that it is not fair.  We have also tested
other specifications of the CCD protocol, for example with secure
communication channels between agents and the TTP, and for the
original definition for the \textsl{abort} sub-protocol: no attack has
been found; but using such channels is not considered as acceptable,
because it requires too much work for the TTP.

Our method, based on the writing of simple state invariants, is of
easy use, and can be implemented in any tool handling agents (or
intruder) knowledge.  It should be very helpful for setting
abstractions for handling unbounded scenarios, and it should very
efficient for bounded verifications, as it has been the case in our
implementation.  We hope that this work will open a highway to the
specification of many other properties, without any more change in the
specification languages and the analysis engines.

\bibliographystyle{abbrv}

\newpage
\tableofcontents

\end{document}